\documentstyle[12pt,epsfig]{article}
\begin{document}
\centerline{\Large\bf Corrections to the Fine Structure Constant}
\centerline{\Large\bf in the Spacetime of a Cosmic String from}
\centerline{\Large\bf the Generalized Uncertainty Principle}
\vspace*{0.050truein}
\centerline{Forough Nasseri\footnote{Email: nasseri@fastmail.fm}}
\centerline{\it Physics Department, Sabzevar University of
Tarbiat Moallem, P.O.Box 397, Sabzevar, Iran}

\begin{center}
(\today)
\end{center}

\begin{abstract}
We calculate the corrections to the fine structure constant
in the spacetime of a cosmic string. These corrections stem from the
generalized uncertainty principle. In the absence of a cosmic string
our result here is in agreement with our previous result.
\end{abstract}

The gravitational properties of cosmic strings are strikingly different
from those of non-relativistic linear distributions of matter. To explain
the origin of the difference, we note that for a static matter
distribution with energy-momentum tensor,
\begin{equation}
\label{1}
T^{\mu}_{\nu}={\rm diag} \left( \rho, -\frac{p_1}{c^2},
-\frac{p_2}{c^2}, -\frac{p_3}{c^2} \right),
\end{equation}
the Newtonian limit of the Einstein equations becomes
\begin{equation}
\label{2}
\nabla^2\Phi=4 \pi G \left( \rho+\frac{p_1+p_2+p_3}{c^2} \right),
\end{equation}
where $\Phi$ is the gravitational potential. For non-relativistic
matter, $p_{\rm i} \ll \rho c^2$ and $\nabla^2 \Phi = 4 \pi G \rho$.
Strings, on the other hand, have a large longitudinal tension. For a
straight string parallel to the z-axis, $p_3=-\rho c^2$, with $p_1$
and $p_2$ vanish when averaged over the string cross-section. Hence,
the right-hand side of (\ref{2}) vanishes, suggesting that straight
strings produce no gravitational force on surrounding matter.
This conclusion is confirmed by a full general-relativistic
analysis. Another feature distinguishing cosmic strings from more
familiar sources is their relativistic motion. As a result, oscillating
loops of string can be strong emitters of gravitational radiation.

The analysis in this letter
is based on thin-string and weak-gravity approximations.
The metric of a static straight string lying along the $z$-axis
in cylindrical coordinates $(t, z, \rho, \phi)$ is given
by\footnote{We use the notation $(t, z, \rho, \phi)$
for cylindrical coordinates and $(t, r, \theta, \phi)$ for spherical
coordinates. Here the mks units have been used.}
\begin{equation}
\label{3}
ds^2=c^2 dt^2-dz^2- (1-h) (d \rho^2+ \rho^2 d \phi^2),
\end{equation}
where $G$ is Newton's gravitational constant, $\mu$ the string mass
per unit length and
\begin{equation}
\label{4}
h=\frac{8G \mu}{c^2} \ln \left( \frac{\rho}{\hat \rho} \right).
\end{equation}
Introducing a new radial coordinate $\rho'$ as
\begin{equation}
\label{5}
(1-h) \rho^2=(1-\frac{8 G \mu}{c^2} ) \rho'^2,
\end{equation}
we obtain to linear order in $\frac{G\mu}{c^2}$,
\begin{equation}
\label{6}
ds^2=c^2 dt^2-dz^2-d \rho'^2-(1-\frac{8 G\mu}{c^2}) \rho'^2 d\phi^2.
\end{equation}
Finally, with a new angular coordinate
\begin{equation}
\label{7}
\phi'=(1-\frac{4G\mu}{c^2})\phi,
\end{equation}
the metric takes a Minkowskian form
\begin{equation}
\label{8}
ds^2=c^2 dt^2-dz^2-d \rho'^2- \rho'^2d \phi'^2.
\end{equation}
So, the geometry around a straight cosmic string is locally identical
to that of flat spacetime. This geometry, however is not globally
Euclidean since the angle $\phi'$ varies in the range
\begin{equation}
\label{9}
0 \leq \phi' < 2 \pi \left( 1- \frac{4G\mu}{c^2} \right).
\end{equation}
Hence, the effect of the string is to introduce an azimuthal `deficit
angle'
\begin{equation}
\label{10}
\Delta=\frac{8\pi G \mu}{c^2},
\end{equation}
implying that a surface of constant $t$ and $z$ has the geometry of a
cone rather than that of a plane \cite{1}.

As shown above, the metric (\ref{6}) can be transformed to a flat metric
(\ref{8}) so there is no gravitational potential in the space outside the
string. But there is a delta-function curvature at the core of the cosmic
string which has a global effect-the deficit angle (\ref 10).

The dimensionless parameter $\frac{G \mu}{c^2}$ plays an important role in the
physics of cosmic strings. In the weak-field approximation
$\frac{G\mu}{c^2} \ll 1$.
The string scenario for galaxy formation requires $\frac{G \mu}{c^2} \sim 10^{-6}$
while observations constrain $\frac{G \mu}{c^2}$
to be less than $10^{-5}$ \cite{1}.

Linet in \cite{2} has shown that the electrostatic
field of a charged particle is distorted by the cosmic string.
For a test charged particle in the presence of a cosmic string
the electrostatic self-force is repulsive and is perpendicular to the
cosmic string lying along the $z$-axis\footnote{Linet
in \cite{2} has used the mks units and in Eqs.(15) and (16) of \cite{2}
has obtained $f^z=f^{\phi}=0$ and
$$
f^{\rho} \sim \left( \frac{2.5}{\pi} \right)
\left( \frac{G \mu}{c^2} \right)
\left( \frac{q^2}{4 \pi \epsilon_0 \rho_0^2} \right)
$$
when $\mu \to 0$. Indeed we can put the fraction $\frac{2.5}{\pi}$
to be approximately equal to $\frac{\pi}{4}$. With this substitution
we obtain (\ref{11}) of this article.}
\begin{equation}
\label{11}
f^{\rho}\simeq\frac{\pi}{4} \frac{G \mu}{c^2} \frac{e^2}{4 \pi \epsilon_0 \rho_0^2},
\end{equation}
where $f^{\rho}$ is the component of the electrostatic self-forcs
along the $\rho$-axis in cylindrical coordinates and $\rho_0$ is the
distance between the electron and the cosmic string.

For the Bohr's atom in the absence of a cosmic string, the electrostatic
force between an electron and a proton is given by Coulomb
law
\begin{equation}
\label{12}
\vec F=\frac{-e^2}{4 \pi \epsilon_0 r^2} \hat{r}.
\end{equation}

As discussed in \cite{3,4,5},
to obtain the fine structure constant in the spacetime of a cosmic
string we assume that the proton located on the cosmic string
lying along the $z$-axis. We also assume that the proton
located in the origin of the cylindrical coordinates and
the electron located at $\rho=\rho_0$, $z=0$ and $\phi=0$.
This means that the electron and the proton are in the plane
orthogonal to the cosmic string.

To calculate the Bohr radius in the spacetime of a cosmic
string we consider a Bohr's atom in the presence of a cosmic string.
For a Bohr's atom in the spacetime of a cosmic string, we
take into account the sum of two forces, i.e.
the electrostatice force for Bohr's atom in the absence of a cosmic
string, given by Eq.(\ref{12}), plus the electrostatic self-force of
the electron in the presence of a cosmic string.
Because we assume that the proton located at the origin of the
cylindrical coordinates and on the cosmic string and also
the plane of electron and proton is perpendicular to the cosmic string
lying along the $z$-axis, the induced electrostatic self-force and the
Coulomb force are at the same direction, i.e. the direction of the
$\rho$-axis in cylindrical coordinates. Therefore, we can sum
these two forces
\begin{equation}
\label{13}
\vec F_{\rm {tot}}= \left( -\frac{e^2}{4 \pi \epsilon_0 \rho_0^2}+\frac{\pi}{4}
\frac{G \mu}{c^2} \frac{e^2}{4 \pi \epsilon_0 \rho_0^2} \right) \hat \rho.
\end{equation}
It can be easily shown that this force has negative value and
is an attractive force ($\frac{\pi G \mu}{4 c^2} < 1$).

The numerical value of Bohr radius in the spacetime of
a cosmic string can be computed by (\ref{13}).
Using Newton's second law, we obtain
\begin{equation}
\label{14}
\frac{m v^2}{\rho_0}= \frac{p^2}{m \rho_0}=
\frac{e^2}{4 \pi \epsilon_0 \rho_0^2}-
\frac{\pi}{4} \frac{G \mu}{c^2} \frac{e^2}{4 \pi \epsilon_0 \rho_0^2},
\end{equation}
where $m$ is the mass of the electron.
Cancelling one $\rho_0$ and rearranging gives
\begin{equation}
\label{15}
p^2=
\frac{me^2}{4 \pi \epsilon_0 \rho_0}
\left( 1 - \frac{\pi}{4} \frac{G \mu}{c^2} \right).
\end{equation}
There is a relationship between the radius and the momentum
\begin{equation}
\label{16}
\rho_{\rm n} p_{\rm n}= n \hbar.
\end{equation}
The product of the radius and the momentum in the left-hand side of
(\ref{16}) is the angular momentum. According to Bohr's hypothesis,
the angular momentum $L$ is quantized in units of $\hbar$.
This means that
\begin{equation}
\label{17}
L_{\rm n}=n \hbar.
\end{equation}
Substituting (\ref{16}) into (\ref{15}) gives
\begin{equation}
\label{18}
\left( \frac{n \hbar}{\rho_{\rm n}} \right)^2 =
\frac{me^2}{4 \pi \epsilon_0 \rho_{\rm n}}
\left( 1 - \frac{\pi}{4} \frac{G \mu}{c^2} \right),
\end{equation}
or
\begin{equation}
\label{19}
\rho_{\rm n}=\frac{4 \pi \epsilon_0 n^2 \hbar^2}{m e^2 \left(
1- \frac{\pi}{4} \frac{G \mu}{c^2} \right)}.
\end{equation}
This equation obtains the radius of the $n^{\rm th}$ Bohr orbit of the
hydrogen atom in the presence of a cosmic string.
In the absence of a cosmic string, the lowest orbit ($n=1$) has a
special name and symbol: the Bohr radius
\begin{equation}
\label{20}
a_B \equiv \frac{4 \pi \epsilon_0 \hbar^2}{m e^2}=
5.29 \times 10^{-11} m.
\end{equation}
Using (\ref{19}),
the Bohr radius ${\hat a}_B$ in the presence of a cosmic string is
\begin{equation}
\label{21}
{\hat a}_B \equiv \frac{ 4 \pi \epsilon_0 \hbar^2}
{me^2 \left( 1-\frac{\pi}{4} \frac{G \mu}{c^2}\right)}.
\end{equation}
From (\ref{20}) and (\ref{21}) we obtain
\begin{equation}
\label{22}
\frac{a_B}{{\hat a}_B}= \left( 1 - \frac{\pi}{4} \frac{G \mu}{c^2} \right).
\end{equation}
In the limit $\mu \to 0$, i.e. in the absence of a cosmic
string, $a_B/{{\hat a}_B} \to 1$.
Inserting $\frac{G\mu}{c^2} \simeq 10^{-6}$ we obtain
\begin{equation}
\label{23}
{\hat a}_B = 
\frac{a_B}{\left( 1- \frac{\pi}{4} \times 10^{-6} \right)}.
\end{equation}
This means that the presence
of a cosmic string causes the value of the Bohr radius 
increases (${\hat a}_B > a_B$). 

Our aim is now to obtain the effective Planck constant
${\hat \hbar}_{\rm eff}$ in the spacetime of a cosmic string
by using the generalized uncertainty principle.
In doing so, we use the modified Bohr radius, ${\hat a}_B$, in the
presence of a cosmic string.

The general form of the generalized uncertainty principle is
\begin{equation}
\label{24}
\Delta x_{\rm i} \geq \frac{\hbar}{\Delta p_{\rm i}} + {\hat\beta}^2 L_{\rm P}^2
\frac{\Delta p_{\rm i}}{\hbar},
\end{equation}
where $\hat{\beta}$ is a dimensionless constant of order one and
$L_P=(\hbar G/c^3)^{1/2}$ is the Planck length.
In the case $\hat{\beta}=0$, (\ref{24}) reads the standard Heisenberg
uncertainty principle
\begin{equation}
\label{25}
\Delta x_{\rm i} \Delta p_{\rm j} \geq \hbar \delta_{\rm {ij}},\;\;\;\;\;\;\;\;{\mbox i,j=1,2,3}.
\end{equation}
There are many derivations of the generalized uncertainty principle,
some heuristic and some more rigorous. Eq.(\ref{24}) can be derived
in the context of string theory and noncommutative quantum mechanics.
The exact value of $\hat{\beta}$ depends on the specific model. The second
term in the right-hand side of (\ref{24}) becomes effective when momentum and
length scales are of the order of the Planck mass and of the Planck length,
respectively. This limit is usually called quantum regime. From
(\ref{24}) we solve for the momentum uncertainty in terms of the
distance uncertainty, which we again take to be the radius of the first
Bohr orbit. Therefore we are led to the following momentum
uncertainty
\begin{equation}
\label{26}
\frac{\Delta p_{\rm i}}{\hbar}= \frac{\Delta x_{\rm i}}{2 \hat{\beta}^2 L_{\rm P}^2}
\left( 1 - \sqrt{1 - \frac{4 \hat{\beta}^2 L_{\rm P}^2}{\Delta x_{\rm i}^2}} \right).
\end{equation}
The maximum uncertainty in the position of an electron in the ground
state in hydrogen atom is equal to the radius of the first Bohr radius,
$a_B$. In the spacetime of a cosmic string, the maximum uncertainty
in the position of an electron in the ground state is equal to the
modified radius of the first Bohr radius, ${\hat a}_B$, see (\ref{21}).

Recalling the standard uncertainty principle
$\Delta x_{\rm i} \Delta p_{\rm i} \geq \hbar$, we define an ``effective'' Planck
constant $\Delta x_{\rm i} \Delta p_{\rm i} \geq \hbar_{\rm eff}$.
From (\ref{24}), we can write
\begin{equation}
\label{27}
\Delta x_{\rm i} \Delta p_{\rm i}\geq
\hbar \left[ 1 + \hat{\beta}^2 L_{\rm P}^2 \left( \frac{\Delta p_{\rm i}}{\hbar} \right)^2
\right].
\end{equation}
So we can generally define the effective Planck constant from
the generalized uncertainty principle
\begin{equation}
\label{28}
\hbar_{\rm eff} \equiv \hbar \left[ 1 + \hat{\beta}^2 L_{\rm P}^2
\left( \frac{\Delta p_{\rm i}}{\hbar} \right)^2 \right].
\end{equation}
Inserting
\begin{equation}
\label{29}
\Delta x_{\rm i}={\hat a}_B=\frac{a_B}{(1- \frac{\pi}{4} \frac{G \mu}{c^2})},
\end{equation}
in (\ref{26}) and using (\ref{28}) give us the effective Planck constant,
${\hat \hbar}_{\rm eff}$, in the spacetime of a cosmic string
\begin{equation}
\label{30}
{\hat \hbar}_{\rm eff}=\hbar \left[ 1 +
\frac{{\hat a}_B^2}{4 {\hat \beta}^2 L_{\rm P}^2}
\left( 1 - \sqrt{1 - \frac{4 {\hat \beta}^2 L_P^2}{{\hat a}_B^2}} \right)^2
\right].
\end{equation}
From (\ref{21}) and $M_{\rm P}=(\hbar c/G)^{1/2}$ which is the Planck mass,
we have
\begin{equation}
\label{31}
\frac{L_{\rm P}}{{\hat a}_B}=
\frac{m e^2 ( 1 - \frac{\pi}{4} \frac{G \mu}{c^2})}
{4 \pi \epsilon_0 M_P^3 G} \ll 1.
\end{equation}
Using $m=9.11 \times 10^{-31} kg$,
$e= 1.6 \times 10^{-19} C$, $c=3.00 \times 10^8 m/s$,
$\epsilon_0=8.85 \times 10^{-12} C^2 N^{-1} m^{-2}$,
$G=6.67 \times 10^{-11} m^3 s^{-2} kg^{-1}$
and $M_P=2.1768 \times 10^{-8} kg$ we obtain
the value of $\frac{L_{\rm P}}{{\hat a}_B} \simeq
\frac{10^{-33}}{10^{-9}} \simeq 10^{-24}$
is much less than one, we can expand (\ref{30}). Therefore we have
\begin{equation}
\label{32}
{\hat \hbar}_{\rm eff} \simeq \hbar \left[ 1
+ {\hat \beta}^2 \left( \frac{m e^2 (1 - \frac{\pi}{4} \frac{G \mu}{c^2})}
{4 \pi \epsilon_0 M_P^3 G} \right)^2 \right].
\end{equation}
So the effect of the generalized uncertainty principle in the presence of
a cosmic string can be taken into account by using
${\hat \hbar}_{\rm eff}$ instead of $\hbar$.
In the absence of a cosmic string, i.e. $\mu \to 0$,
Eq.(\ref{32}) leads us to our previous result in \cite{6}.
In Ref.\cite{3}, we obtained the fine structure constant, ${\hat \alpha}$,
in the spacetime of a cosmic string
\begin{equation}
\label{33}
{\hat \alpha}=\alpha \left( 1 - \frac{\pi}{4} \frac{G \mu}{c^2} \right),
\end{equation}
where $\alpha$ is the fine structure constant,
$\alpha= \frac{e^2}{4 \pi \epsilon_0 \hbar c}$.
Substituting the effective Planck constant ${\hat \hbar}_{\rm eff}$
from (\ref{32}) into (\ref{33}) we obtain the effective and corrected
fine structure constant in the presence of a cosmic string by using
the generalized uncertainty principle
\begin{equation}
\label{34}
{\hat \alpha}_{\rm eff}=
\frac{e^2}{4 \pi \epsilon_0 {\hat \hbar}_{\rm eff} c}
\left( 1 - \frac{\pi}{4} \frac{G \mu}{c^2} \right).
\end{equation}
From (\ref{32}) and (\ref{34}) one can obtain 
\begin{equation}
\label{35}
{\hat \alpha}_{\rm eff} \simeq \left( \frac{e^2}{4 \pi \epsilon_0 \hbar c}
-\frac{\pi}{4} \frac{G \mu}{c^2} \frac{e^2}{4 \pi \epsilon_0 \hbar c} \right) \left[ 1
- {\hat \beta}^2 \left( \frac{m e^2 ( 1 - \frac{\pi}{4} \frac{G \mu}{c^2})}
{4 \pi \epsilon_0 M_P^3 G} \right)^2 \right].
\end{equation}
This equation can be rewritten
\begin{eqnarray}
\label{36}
{\hat \alpha}_{\rm eff} & \simeq & \left( \frac{e^2}{4 \pi \epsilon_0 \hbar c}
-\frac{\pi}{4} \frac{G \mu}{c^2} \frac{e^2}{4 \pi \epsilon_0 \hbar c} \right) \nonumber\\
& \times & \left[ 1
- {\hat \beta}^2 \times 9.30 \times 10^{-50} \left(1- 2 \times \frac{\pi}{4}
\times 10^{-6} \right) \right],
\end{eqnarray}
where we used $(1- \frac{\pi}{4} \frac{G \mu}{c^2} )^2 \simeq
(1-2 \times \frac{\pi}{4} \frac{G \mu}{c^2} )$ and
$\frac{G \mu}{c^2} \sim 10^{-6}$.
From (\ref{33}) and (\ref{36}) we conclude
\begin{equation}
\label{37}
{\hat \alpha}_{\rm eff} \simeq {\hat \alpha} \left[ 1
- {\hat \beta}^2 \times 9.30 \times 10^{-50} \left( 1- 2 \times \frac{\pi}{4} \times
10^{-6} \right) \right].
\end{equation}
This equation shows the corrections to the fine structure constant
in the spacetime of a cosmic string from the generalized uncertainty
principle. In the absence of a cosmic string the expression inside
the parenthesis in the right-hand side of (\ref{37}) is equal to one
and we are led to our previous result in \cite{6}.
In other words, in the absence of a cosmic string our result here,
given by (\ref{37}), is in agreement with
our previous result in \cite{6}.\\
{\bf Acknowledgments}: I thank Amir and Shahrokh for useful helps.

\end{document}